\documentclass[final,3p,times,twocolumn]{elsarticle}

\usepackage{float}
\journal{Physics Letters A}

\usepackage{graphicx}
 \usepackage{epsfig}
\begin{document}
\def\ds{\displaystyle}
\begin{frontmatter}

\title{Self-acceleration in non-Hermitian Systems}
\author{C. Yuce$^{\dagger}$, Z. Turker$^{\star}$}
\address{Department of Physics, Anadolu University, Turkey$^{\dagger}$, \\Faculty of Engineering, Near East University, Nicosia, Cyprus$^{\star}$}
\ead{$^{\dagger}$cyuce@anadolu.edu.tr} \fntext[label2]{}
\begin{abstract}
We study self-acceleration in PT and non-PT symmetric systems. We find some novel wave effects that appear uniquely in non-Hermitian systems. We show that integrable self-accelerating waves exist if the Hamiltonian is non-Hermitian. We find that self-accelerating constant intensity waves are possible even when gain and loss are not balanced in the system. 
\end{abstract}


\end{frontmatter}


\section{Introduction}
non-normalizable
A self-accelerating wave packet is an accelerating wave packet whose acceleration is not equal to the one obtained by the Ehrenfest theorem. Therefore a self-accelerating wave packet must be non-integrable (non-normalizable), otherwise the Ehrenfest theorem is satisfied. However, this does not necessarily mean that any non-integrable wave packet is a self-accelerating wave packet. One has to perform analytical calculations to find whether a given non-normalizable wave packet self-accelerates or not. More than three decades ago, Berry and Balasz theoretically showed that the exact solution of the Schrodinger equation with a linear potential is also an exact self-accelerating nondiffracting solution for the free particle Schrodinger equation \cite{berry}. In other words, the Airy wave packet moves with a constant acceleration in the absence of any force. For practical applications, non-normalizable Airy wave function must be truncated. In the last decade, truncated Airy wave packets were routinely realized by many groups around the world \cite{deney1,deney2,deney3,deney4,deney5,ek}. These experiments have been performed within the context of optics. The key idea here is that the paraxial equation of diffraction in optics and the Schrodinger equation in quantum mechanics are equivalent. Note that truncated Airy wave, which is necessary for an experimental realization breaks self-acceleration since any integrable (normalizable) wave satisfies the Ehrenfest's theorem. However, its non-diffracting and accelerating feature are not lost within a large distance. In fact, the main lobe of the truncated Airy waves was found to follow a parabolic trajectory. The accelerating Airy wave was also obtained for free electrons using a nanoscale hologram
technique \cite{electron}. Airy wave is the first self-accelerating wave appeared in the literature. Many researchers have explored other types of accelerating beams such as parabolic cylinder waves \cite{cemplayeni}, Weber and Mathieu beams \cite{matweb1,matweb2,matweb3} and accelerating beams in curved space \cite{curvedMax}. We stress that not only the Schrodinger equation but also other equations known in physics such as nonlinear paraxial \cite{nonlin1,nonlin2,nonlin3,nonlin4} and nonparaxial equation \cite{nonparax1,nonparax2} admit self-accelerating solution. Self-accelerating waves were also analyzed in nonuniform systems such as matter waves in an expulsive potential \cite{cemmodern} and periodical photonics systems \cite{onemli01,onemli02,onemli03,102,9,9999,10,101}. \\
Another interesting systems that have attracted great attention  in the last two decades are the systems governed by non-Hermitian Hamiltonians with $\mathcal{PT}$ symmetry, where $\mathcal{P}$ and $\mathcal{T}$ operators are parity and time reversal operators, respectively. It is well known that spectrum of a $\mathcal{PT}$ symmetric non-Hermitian Hamiltonian is real unless non-Hermitian degree exceeds a critical number. If it exceeds the critical number, $\mathcal{PT}$ symmetry is spontaneously broken and the energy spectrum becomes either partially or completely complex. We note that there exist non-$\mathcal{PT}$ symmetric systems with all-real spectra \cite{nonPT1,nonPT2}. Non-Hermitian $\mathcal{PT}$ symmetric systems are not just of theoretical interest. The first experiment on a $\mathcal{PT}$ symmetric system with balanced gain and loss was realized in an optical system \cite{deney34}. \\
Novel wave effects have been shown to arise in non-Hermitian systems. These are, for example, unidirectional light transport \cite{uni1,uni2}, interesting  topological insulating effects \cite{cemo2,cemo3}, single-mode lasers \cite{las1,las2,las3}, Majorana edge modes \cite{cemo1}, potential violation of the no-signalling principle \cite{nosignal} and constant intensity waves \cite{ConsIntWaves}. In this paper, we will explore some other novel wave effects that are not possible in Hermitian systems. These are the self-accelerating constant intensity waves and normalizable self-accelerating waves. 

\section{Self-accelerating Solution}

We start with the following Hamiltonian operator with a complex potential
\begin{equation}\label{cn10}
H=-\frac{1}{2}\frac{\partial^2}{\partial x^2}+U,~~~~~U=~V_R+i ~V_I
\end{equation}
where $V_R(x,t) $ and $V_I (x,t) $ are the real and imaginary parts of the time dependent complex potential, respectively. The Hamiltonian operator governs the time evolution of the complex valued wave function $\Psi(x,t)$ in the form $\ds{H\Psi=i\frac{{\partial}\Psi}{\partial  t}}$.\\
This Hamiltonian is $\mathcal{PT}$ symmetric if $\ds{V_R(-x,-t)=V_R(x,t)}$ and $\ds{V_I(-x,-t)=-V_I(x,t)}$. We note that the $\mathcal{PT}$ symmetry of the Hamiltonian doesn't guarantee the reality of the spectrum. If the $\mathcal{PT}$ symmetry is spontaneously broken, i.e., $\ds{\mathcal{PT}\Psi\neq\mp\Psi}$, then the system admits complex energy eigenvalues. Generally speaking, neither Hermiticity nor $\mathcal{PT}$ symmetry serves as a sufficient condition for the reality of the spectrum. In this paper, we will not study the reality of the spectrum of the above Hamiltonian. Instead, we will investigate self-acceleration in our non-Hermitian system by solving the corresponding Schrodinger equation exactly. We will explore novel effects that arises only when the Hamiltonian is non-Hermitian. \\
An exact analytical self-accelerating solution has so far obtained only for two Hermitian potentials in the literature: These are the free particle potential \cite{berry} and inverted harmonic potential, $V_R=-x^2, V_I=0$ \cite{cemplayeni}. Our aim is to obtain exact self-accelerating solutions in the non-Hermitian case. Let us first rewrite the Schrodinger equation in the accelerating frame by transforming the coordinate according to $\ds{q=x-x_c}$, where the time-dependent function $\ds{x_c(t)=\frac{1}{2} a t^2}$ describes translation and the constant $a$ is the acceleration. Under this coordinate transformation, the time-derivative operator transforms as $\ds{\partial_t\rightarrow\partial_t-\dot{x_c }  \partial_{q}}$, where dot denotes derivation with respect to time. In the accelerating frame, we will seek
the solution of the form 
\begin{equation}\label{s378ts}
\Psi(q,t)=\exp{\left(i\dot{x_c}{q}+i  \int \frac{G}{\psi^2} dq^{\prime} +iS(t) \right)}~\psi({q}) 
\end{equation}
where $\ds{\dot{S}=\frac{1}{2} \dot{x_c}^2}-\mu$, the constant $\mu$ is a real number and $\ds{\psi(q)}$ and $\ds{G(q)}$ are real valued functions to be determined below. \\
Substituting the transformation (\ref{s378ts}) into the Schrodinger equation in the accelerating frame yields equations for $G(q)$ and imaginary part of the potential $V_I$. They are given by
\begin{eqnarray}\label{sutfbcp}
G^2&=&\psi^3(\psi^{\prime\prime}+2(\mu-a q-V_R)\psi) \\
V_I&=&\frac{G^{\prime}}{2\psi^2} \label{sutfbcp2}
\end{eqnarray}
where prime denotes derivation with respect to $q$. Here we assume that $V_R$ and $V_I$ depend only on $q$.\\
These are the equations to be satisfied by $G$ and $\psi$ in the accelerating frame. For given $V_R$ and $V_I$, one can solve these equations to find $G$ and $\psi$. Below, we will analyze these two equations in detail.\\
Let us first review the self-accelerating waves in the Hermitian limit $\ds{V_I=0}$. In this case, the Ehrenfest's theorem always works for normalizable wave packets. Therefore, a self-accelerating solution can only be possible for a nonintegrable wave packet. An exact self-accelerating solution is available for the free particle $V_R=0$ \cite{berry}. This solution can be recovered by solving the equation (\ref{sutfbcp}) with $\ds{G=0}$.  It is given by $\ds{\psi(q)=Ai\left((2a)^{1/3}(q-\mu/a)\right)}$. The Airy wave function is not square integrable and does not obey the Ehrenfest's theorem since the center of mass of the non-normalizable Airy function cannot be defined. This Airy wave is a stationary solution in the accelerating frame. Transforming backward yields the accelerating solution in the lab. frame. One can see that the constant 
$a$ is practically determined by the width of the Airy function. Therefore acceleration of the Airy wave can be changed just by varying the initial width of the Airy function.\\
Having obtained the standard self-accelerating solution in the Hermitian limit, we can now study the non-Hermitian case. The novel wave effect that arises uniquely in non-Hermitian systems are the existence of constant intensity self-accelerating waves and integrable self-accelerating waves. Below, we will investigate them separately.

\subsection{Self-accelerating constant intensity waves}

The plane wave for free particle is the well known constant intensity wave in physics. If a plane wave is scattered from a potential, then its intensity varies with position at a given time. Obtaining constant intensity waves in a system with potential is important. Recently, the concept of constant intensity waves was extended to non-Hermitian potentials \cite{ConsIntWaves}. So far, constant intensity waves have been analyzed by considering that they move with a constant velocity. A question arises? Is it possible to realize a constant intensity wave which speeds up with a constant self-acceleration? In fact, they can not be realizable in Hermitian systems. As will be shown below, non-Hermitian systems can support such waves.\\
Let us set $\ds{\psi(q)=1}$ to study self-accelerating waves that have constant intensities. The equations (\ref{sutfbcp},\ref{sutfbcp2}) yield the condition for the imaginary part of the potential to observe such waves. It is given by $\ds{V_I=-\frac{a+V_R^{\prime}}{2\sqrt{2(\mu-aq-V_R)}}}$. Here the constant $\ds{\mu}$ must be chosen in such a way that the inside of the square root should be positive for all $q$. As an example, consider the following inverted harmonic potential: $\ds{V_R=-V_0^2q^2}$, where $V_0$ is a real number \cite{inverted}. Therefore the imaginary part of the potential reads $\ds{V_I=-\frac{a-2V_0^2q}{2\sqrt{2(\mu-aq+V_0^2q^2)}}}$. We stress that self-acceleration breaks the $\mathcal{PT}$ symmetry for this potential. In other words, this potential is $\mathcal{PT}$ symmetric only when $a=0$. Furthermore, gain and loss are not balanced in the system, i.e., $\ds{\int_{-\infty}^{\infty}V_Idx=-\frac{a}{\sqrt{2}|V_0|}\neq0}$ for $\ds{\mu{>}\frac{a^2}{4V_0^2}}$. One can see that the denominator is always positive unless $\ds{\mu}$ is smaller than $\ds{\frac{a^2}{4V_0^2}}$. So, we say that constant intensity wave with the continuous parameter $\mu$ is possible for this complex potential. The case with $\ds{\mu=\frac{a^2}{4V_0^2}}$ is interesting since the imaginary part of the potential is constant: $\ds{V_I=V_0/\sqrt{2}}$ (Note that $\ds{G=\sqrt{2} V_0 (q-\frac{a}{2V_0})}$). This implies that only gain (loss) is present in the system when $\ds{V_0>0}$ ($\ds{V_0<0}$). This result is striking since the existence of accelerating wave packet with constant intensity is supported even though either gain or loss exists in the system. This striking observation is, up to the best of our knowledge, the first example of a non-growing/non-decaying wave packet in a quantum system with either gain or loss. Normally, we expect that a wave packet grows/decays in the presence of gain/loss. For a physical realization of such a constant intensity wave packet, one has to truncate the wave packet since they carry infinite energy. As noted in \cite{ConsIntWaves}, the wave packet retains many of their exciting properties when being truncated . \\
As a second special example, suppose $\ds{V_R=-aq-V_0^2q^n}$, where $n=2,4,6,...$. In the absence of gain and loss, $V_I=0$, then any normalizable wave packet in such a potential accelerates hyperbolically according to the Ehrenfest's theorem. In the presence of the gain and loss, the Ehrenfest's theorem fails and our shape invariant accelerating constant intensity wave solution arises.  If we solve the equations (\ref{sutfbcp},\ref{sutfbcp2}) for this potential at $\mu=0$, we obtain that $\ds{V_I= -\frac{n}{2\sqrt{2}}V_0 q^{\frac{n}{2}-1}}$. \\
To this end, we would like to mention that generalization of self-accelerating constant intensity waves to nonlinear domain is straightforward. Let us replace $\ds{V_R{\rightarrow}V_R+\sigma|\Psi|^p}$ in the equation (\ref{cn10}), where the constant $\ds{p}$ is the degree of the nonlinearity ($\ds{p\neq0}$) and  $\sigma$ is the strength of nonlinear interaction. If, for example, $p=2$, then the equation (1) becomes 1-D nonlinear Schrodinger (NLS) equation with complex potential. Apparently, our nondiffracting constant intensity waves are also solutions of the generalized nonlinear equation. The self-accelerating solution for the nonlinear system can be obtained if we replace $\ds{\mu\rightarrow\mu+2\sigma}$ in the solution of the linear system. We emphasize the remarkable fact that two different systems with different values of $p$ have the same solutions up to a constant phase. The degree of the nonlinear interaction term has nothing to do with the intensity of the propagating wave.
\begin{figure}[t]\label{20}
\includegraphics[width=9cm]{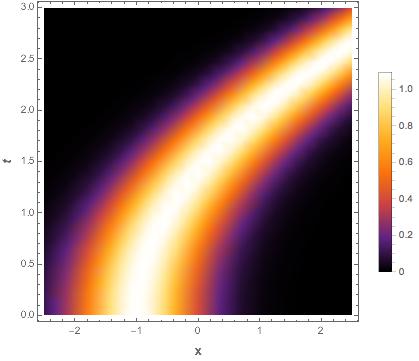}
\caption{ The density plot shows the propagation of the Gaussian wave in the purely imaginary potential, $\ds{V_I=- \omega^2 q^2-aq+\omega/2}$, where $\omega=a=1$ and $q=x-x_c$. This localized wave packet is bended as a result of the self-acceleration. }
\end{figure}

\subsection{Integrable self-accelerating waves} 

In an Hermitian system, self-acceleration can not be observed for normalizable wave packets as a result of the Ehrenfest's theorem. There is no general theory that accounts for the acceleration of non-normalizable wave packets. The fact that self-acceleration is only possible for such wave packets is not good from the experimental point of view since non-normalizable wave packets can not be realizable in practice. Therefore, self-accelerating waves must be truncated, which in turn implies that the Ehrenfest's theorem is satisfied. As a result, no true self-acceleration can be observed in practice. Instead, one can observe interesting accelerating behavior. As an example, it was observed that main lobe of self-accelerating truncated Airy wave moves on a curved trajectory \cite{deney1}. \\
In this paper, we show that normalizable self-accelerating waves are possible if the system is non-Hermitian. As opposed to non-normalizable self-accelerating wave packets, truncation is not necessary in our system. So, true self-acceleration is possible in practice for non-Hermitian systems. \\
Below we explore self-accelerating integrable waves for free particle with gain and loss, i.e., $V_R=0$ but $V_I\neq0$. To obtain integrable self-accelerating waves, we solve the equations (\ref{sutfbcp},\ref{sutfbcp2}) for the boundary conditions $\ds{\psi(x\rightarrow \mp \infty)=0}$. An interesting point to observe is that such a solution doesn't exist for Hermitian potentials, since for $\ds{G=0}$ ($\ds{V_I=0}$) exact wave function is the Airy function, which is not integrable. Only when $\ds{G\neq0}$, i.e., $\ds{V_I\neq0}$, integrable wave solution may exist. Therefore, integrable solutions are a direct consequence of the non-Hermitian nature of the system. \\
To illustrate the properties of such normalizable self-accelerating solutions, we consider the following specific example, $\ds{V_R=0}$ and $\ds{V_I=- \omega^2 q^2-aq+c}$, where $\omega$ and $c$ are constants. Apparently this potential is not $\mathcal{P} \mathcal{T}$ symmetric. Additionally, the gain and loss in the system are not balanced. If $\ds{c=0}$ for $a>0$, then only loss is present in the system. If $\ds{c>0}$ for $a>0$, the system is dominantly lossy with a small interval of gain region. This system has an exact self-accelerating Gaussian wave solution when $\ds{c=\omega/2}$: $\ds{\psi=\exp{(-\frac{\omega}{2} q^2- \frac{a}{\omega}q)}}$ with $\ds{2\mu=\omega-a^2/\omega^2}$ according to (\ref{sutfbcp},\ref{sutfbcp2}). Note also that $\ds{G={\omega} ~q~ \psi^2}$. This wave packet, which is integrable when $\ds{\omega>0}$, accelerates with the acceleration $\ds{a}$ in the lab. frame. But this acceleration is not due to real force but gain and loss in the system. Normally, one expects that the gain and loss in the system leads to either growth or reduction of particle number. However, in our case, it causes the wave packet to accelerate. Note also that the gain and loss lead to localization of free particle as well as self-acceleration. Recall that the real part of the  potential is zero, which implies that there exists no trapping potential. As a limiting case, if no gain and loss are present in the system, i.e., $\omega=a=0$, then the wave packet becomes extended. In the Fig-1, we plot the propagation of such a Gaussian wave in the lab frame. As can be seen from the figure, the beam has curved trajectory. We stress that self-acceleration can be observed no matter how long it moves since the self-accelerating wave packet doesn't have to be truncated.

\subsection{Nonlinearity Mimicking in non-Hermitian Systems}

Having discussed self-accelerating constant intensity waves and integrable waves seperately, let us now study another interesting solution. Suppose first that $\ds{G=\sqrt{2}\sigma\psi^n}$, where $\ds{\sigma}$ and $n>0$ are real valued constants. Then the real valued wave function $\psi$ satisfies a nonlinear equation $\ds{-\psi^{\prime\prime}+2(\mu-a q-V_R)\psi+2\sigma^2\psi^{2n-3}=0}$ according to (\ref{sutfbcp}). We emphasize that our original system is not a true nonlinear system. Instead our self-accelerating solution in the transformed coordinate system coincides with the solution of the above nonlinear equation. As a special case, this equation becomes the well known nonlinear Schrodinger equation when $\ds{n=3}$. As an example, let us study accelerating dark soliton solution by considering $\ds{V_R=-aq}$ when $n=3$. The dark soliton solution reads $\ds{\psi=\tanh(\sigma~q)}$ with $\ds{\mu=-\sigma^2}$. The corresponding density is an anti-symmetric function of $q$ and equals to zero at $q=0$. Furthermore the dark solution appears in the uniform background since the density goes to $1$ at $q=\mp \infty$. In the lab. frame, this solitonic wave packet maintain a constant shape and size as they propagate at a constant acceleration. Note that the corresponding imaginary part of the potential is given by $\ds{V_I=3/\sqrt{2}~\sigma^{2}sech^2(\sigma~q) }$, which is equal to $3/\sqrt{2} \sigma^2$ at $q=0$ and decreases as $|q|$ increases and becomes practically zero at large values of $q$. As can be seen easily, $\ds{V_I>0}$ at any value of $q$ and vice versa. This means that the system has either gain or loss depending on the sign of $\sigma$. In other words, the system is neither a $\mathcal{P} \mathcal{T}$ symmetric system nor a gain/loss balanced system. We note that the wave function vanishes when the imaginary part of the potential is absent, $\sigma=0$. Therefore we say that this nontirivial solution is a consequence of the non-Hermitian character of the system. Finally we would like to mention that our system is fully stable since nonlinearity comes to our linear system effectively.\\
To sum up, we have analyzed self-accelerating solutions in non-Hermitian systems. We have predicted the existence self-accelerating constant intensity waves and normalizable waves, which are impossible to realize in Hermitian systems. We have found that self-accelerating constant intensity waves are possible even when gain and loss are not balanced. We have also discussed that nonlinearity can be used to mimic a non-Hermitian system.

\end{document}